\documentclass[11pt,a4paper]{article}

\usepackage{epic,eepic,epsfig,amsmath,amssymb,amsthm}
\usepackage{t1enc}

 \usepackage[francais]{babel}

\usepackage{color}

\usepackage{bbm}

\def\virgp{\raise 2pt\hbox{,}}

\renewcommand{\geq}{\geqslant}
\renewcommand{\leq}{\leqslant}
\def\N{{\mathbb N}}

\def\R{{\mathbb R}}

\def\virgp{\raise 2pt\hbox{,}}
\def\cdotpv{\raise 2pt\hbox{;}}

\def\1{\mathbbm{1}}

\theoremstyle{remark}

\theoremstyle{definition}

\theoremstyle{definition}

\theoremstyle{definition}

\begin{document}

\title{Structural stability of Lattice Boltzmann schemes}

\author{Claire David $^\dag$, Pierre Sagaut $^\ddag$}

\maketitle
\centerline{$^\dag$ Universit\'e Pierre et Marie Curie-Paris 6}
\centerline{Laboratoire Jacques Louis Lions - UMR 7598}
\centerline{Boîte courrier 187, 4 place Jussieu, F-75252 Paris
cedex 05, France}

\vskip 0.5cm

\centerline{$ ^\ddag$ Universit\'e Aix-Marseille}
\centerline{ST JEROME,
Avenue Escadrille Normandie Niemen}
 \centerline{
163 Avenue de Luminy,
case 901
13009 Marseille, France}


\maketitle

\begin{abstract}
The goal of this work is to determine classes of traveling
solitary wave solutions for Lattice Boltzmann schemes by means of an hyperbolic ansatz. It is
shown that spurious solitary waves can occur in finite-difference
solutions of nonlinear wave equation. The occurence of such a
spurious solitary wave, which exhibits a very long life time,
results in a non-vanishing numerical error for arbitrary time in
unbounded numerical domain. Such a behavior is referred here to
have a structural instability of the scheme, since the space of
solutions spanned by the numerical scheme encompasses types of
solutions (solitary waves in the present case) that are not
solutions of the original continuous equations. This paper extends
our previous work about classical schemes to Lattice Boltzmann schemes (\cite{ClDavid1}, \cite{ClDavid2}, \cite{ClDavid3},\cite{ClDavid4}).
\end{abstract}

\vspace{0.5cm}
\noindent \textbf{Key Words}:\\
\noindent Lattice Boltzmann schemes; solitary waves; numerical flux; structural stability.\\ \\


\section{Introduction:}
\label{DRP}

\noindent The lattice Boltzmann method (\emph{LBM}) is used for the numerical simulation of physical phenomena, and serves as an alternative to classical solvers of partial differential equations. The primary domain of application is fluid dynamics; it is specially used to obtain the numerical solution of the incompressible, time-dependent Navier-Stokes equation.\\

\noindent The strength of the Lattice Boltzmann method is due to its ability to easily represent complex physical phenomena, ranging from multiphase flows to fluids with chemical reactions. The principle is
to "mimic" at a discrete level the dynamics of the Boltzmann equation. Since it is based on a molecular description of a fluid, the knowledge of the microscopic physics can directly be used to formulate the best fitted numerical model.\\

\noindent This method can be regarded as either an extension of the lattice gas automaton (\emph{LGA}) \cite{Hardy}, \cite{Humieres1},\cite{Frisch}, or a special discrete
form of the Boltzmann equation from kinetic theory. Although the connection
between the gas kinetic theory and hydrodynamics has long been established,
the Lattice Boltzmann method (\emph{LBM}) needs additional special discretization of velocity space to
recover the correct hydrodynamics. Due to the very same reason, the \emph{LBM} works exactly
opposite traditional \emph{CFD} methods in deriving working schemes: \emph{LBM} uses Navier-Stokes equations as its target while traditional \emph{CFD} methods use Navier-Stokes equations
as their starting point.\\

\noindent The \emph{LBM} is applicable to the isothermal flow regime, i.e., the weakly compressible,
low-Mach-number limit. This flow regime is traditionally treated as "incompressible", although there are \emph{CFD} methods constructed to compute the Navier-Stokes equations in
this regime. The argument for treating very low-Mach-number flows as incompressible
is pragmatic rather than physical. The direct calculation of the isothermal Navier-Stokes
equations requires time steps sufficiently small to resolve acoustic waves across a computational
cell. This time step may be vastly smaller than the time scales of interest for the bulk
fluid motion. Thus the computational cost of the many additional time steps required by an
isothermal calculation may be vastly higher than the cost of an incompressible calculation.
Of course, in reality there is no fluid or flow that is absolutely incompressible (i.e., with
infinite acoustic velocity).

\section{The Lattice Boltzmann method}

\noindent The lattice Boltzmann (\emph{LB}) method follows the same idea as its predecessor the Lattice Gas Automata (\emph{LGA}) when
it also considers the fluid on a lattice with space and time discrete. Instead of directly describing the fluid by discrete particles and, thus Boolean variables,
it describes the fictitious system in terms of the probabilities of presence of the
fluid particles. A lattice Boltzmann numerical model simulates the time and space evolution of kinetic quantities, the particle
distribution functions $f_j(\vec{r}, t)$, \mbox{$0 \leq j \leq J$}, \mbox{$J\,\in\,\N^\star$}.\\

\noindent The lattice Boltzmann equation is obtained by ensemble averaging
the equation

\begin{equation}
\label{Boltzmann}
\langle N_j\left (\vec{r}+\Delta \,t\,\vec{v}_j,t+\Delta \,t \right )\rangle=\langle N_j\left (\vec{r},t \right ) \rangle+\langle \Omega_j\left (N \right )\rangle
\end{equation}

\noindent where $\langle N_j\left (\vec{r},t \right ) \rangle$ denotes the average number of particles at space position $\vec{r}$ and time $t$.\\
\noindent The system is supposed to satisfy the Boltzmann molecular chaos hypothesis,
i.e. the fact that there is no correlation between particles entering a collision. Thus, the
collision operator can be expressed as $\langle \Omega_i\left (N \right )\rangle= \Omega_i\langle N \rangle$, which leads to the \emph{Lattice Boltzmann equation}:

\begin{equation}
f_j\left (r+\Delta \,t\,v_j,t+\Delta \,t\right )=f_j\left (r,t \right ) + \Omega_j\left (f \right )
\end{equation}

\noindent where, for \mbox{$j\,\in\,\N$}, $f_j = \langle N_j\rangle$ denotes the probability to have a fictitious fluid particle of
velocity $v_j$ entering lattice site $\vec{r}$ at time $t$. The $f_j$ are also called the \emph{fluid fields}, or
the \emph{particle distribution functions}.\\

   The collision operator is normally a non-linear expression and requires a lot of computation time \cite{Chopard1}.
  In a big lattice, e.g. 3D model, the computation becomes impossible even on
 a massively parallel computer. To overcome this problem, Higuera et al. \cite{Higuera1}, \cite{Higuera2}
  proposed to linearize the collision operator around its local equilibrium solution to
  reduce the complexity of the operation. Using this idea, Bhatnager, Gross and Krook introduced the \emph{BGK} \emph{lattice}
  (\emph{LBGK}) \cite{Bhatnagar}, in which the collision between particles is described
  in terms of the relaxation towards a local equilibrium distribution. The \emph{LBGK}
  is considered to be one of the simplest forms of the Lattice Boltzmann equation
  and is mathematically expressed as

\begin{equation}
\label{LBeq}f_j (\vec{r}+ \vec{e}_j\,t+\Delta t)=f_j (\vec{r} ,t )-
\displaystyle \frac{1}{\tau}\, \left \lbrace f_j (\vec{r} ,t )-f_j^{eq}(\vec{r} ,t )\right \rbrace  j\,\in \,\lbrace 0,\,\hdots,\,J\rbrace ,\,  J \,\in\,\N^\star
\end{equation}

  \noindent where $\tau$ is the single relaxation time, which is a free parameter of the model to determine
  the fluid viscosity, and $f_j^{eq}$, \mbox{$j0,\hdots,J$}, denote the local equilibrium functions, which are
  functions of the density and the flow velocity $\vec{u}$.\\

\noindent In the lattice Boltzmann method, the space variable vacetor $\vec{r}$ is supposed to live in a lattice $\cal L$ included in an
Euclidian space of dimension $d$, $d \geq 1$.\\
\noindent The velocity belongs to a finite set $\cal V$ composed by
given velocities $\vec{e}_j$, $ j\,\in \,\lbrace 0,\,\hdots,\,J\rbrace$, $J \,\in\,\N^\star$, chosen in such a way that

\begin{equation}
\vec{r}\,\in\,{\cal L}\,\, \text{and} \, \vec{e}_j\,\in\,{\cal V }\,\Rightarrow\,\vec{r}+\Delta\,t \,\vec{e}_j\,\in\,\cal L
\end{equation}
\noindent where $\Delta \,t$ denotes the time step.\\

\noindent The set of velocities $\cal V$ is supposed to be invariant by space reflection, i.e.:

\begin{equation}
\vec{e}_j\,\in\,{\cal V}\, \Rightarrow\, \exists\,l\,\in\,{\cal V}\,:\,\vec{e}_l=- \vec{e}_j\,\in\,\cal V
\end{equation}

\noindent The numerical scheme is thus defined through the evolution of the population $f_j(\vec{r}, t)$, with
$\vec{r}\,\in\,\cal L$ and $ j\,\in\,\lbrace 0,\,\hdots,\,J\rbrace$ towards a distribution $f_j(\vec{r}, t + \Delta \,t)$ at a new discrete time.\\
\noindent The scheme has two steps that take into account successively the left and
right hand sides of the Boltzmann equation (\ref{Boltzmann}). The first step describes the relaxation
of particle distribution towards the equilibrium. It is local in space and
nonlinear in general.  The second step is the collision process. Both steps have to be computed separately.


\noindent In general \emph{LB} model, denoted by $D_d\,Q(J + 1)$, \mbox{$J\,\in\,\N^\star$}, the actual macroscopic density, which is function of the space vector $\vec{r}$, and the time variable $t$,
is obtained as a summation of the microscopic particle distribution functions:

\begin{equation}
\rho(\vec{r},t)=\displaystyle \sum_{j=0}^J  f_j(\vec{r},t)
\end{equation}

\noindent The macroscopic velocity is calculated as the average of the microscopic velocities $\vec{e}_j$, $j=0, \hdots,J$, weighted by the related distribution functions:

\begin{equation}
\vec{u}(\vec{r},t)=\displaystyle \frac{1}{\rho(\vec{r},t)}\, \displaystyle \sum_{j=0}^J  c\, f_j(\vec{r},t)\,\vec{e}_j
\end{equation}

\noindent In each time step, in  each  node, particles are streamed on to the neighbouring
nodes, which thus lead to new distribution functions $f_j^{\ast}$:

\begin{equation}
\label{Streaming}
  f_j(\vec{r}+\vec{e}_i\,\Delta t,t+\Delta t) =f_j^\ast(\vec{r},t+\Delta t)   \quad, \quad j=0,\hdots, J 
\end{equation}

\noindent Also, once  in  each  time  step,  the  particles  in  each  node  collide,  which
is modeled as a relaxation of the distribution functions towards the
equilibrium distributions

\begin{equation}
\label{Collision}
f_j^\ast(\vec{r},t+\Delta t) = f_j(\vec{r},t) +\displaystyle  \frac{1}{\tau } \left (f_j^{eq}(\vec{r},t)-f_j(\vec{r},t)\right)
\end{equation}

\noindent with:

$$f_j^{eq}(\vec{r},t)=\rho(\vec{r},t) \,w_j\,\left [1+ \displaystyle \frac{ \vec{e}_j \cdot \vec{u}(\vec{r},t)}{c_s^2}+
\displaystyle \frac{\left ( \vec{e}_j \cdot \vec{u}(\vec{r},t)\right)^2}{c_s^4}
-
\displaystyle \frac{\left ( \vec{u}(\vec{r},t)  \cdot \vec{u}(\vec{r},t)\right)^2}{2\,c_s^2}
\right]  \quad, \quad j=0,\hdots, J$$

\noindent   where $c_s$ is the lattice sound speed, such that:

$$c_s^2=\displaystyle \sum_{j=0}^J w_j\,{\vec{e}_j}^{\,\,2} $$

\noindent  and where the weights $w_j$, $j=0,\hdots, J$, are determined by the velocity set. Of course, one has:

$$\displaystyle \sum_{j=0}^J w_j=1$$

The lattice Boltzmann equation takes both steps into account:

$$ f_j(\vec{r}+\vec{e}_i\,\Delta t,t+\Delta t) =f_j^\ast(\vec{r},t+\Delta t) 
= f_j(\vec{r},t) +\displaystyle  \frac{1}{\tau } \left (f_j^{eq}(\vec{r},t)-f_j(\vec{r},t)\right)  \quad, \quad j=0,\hdots, J$$

\noindent The right-hand  side  gives thus the  distribution  of  particles  once  collisions have occured,
while the left-hand side represents particles appearing in neighbouring nodes in the next time step.\\

\noindent For our study, it is important to take into account the way the related algorithm is implemented:\\

\begin{enumerate}
\item[\emph{i}.] Initialization step of $\rho$, $\vec{u}$, and the microscopic densities $f_j$, $f_j^{eq}$, $j=0,\hdots, J$.\\
 
\item[\emph{ii}.] Streaming step: densities $f_j$ are moved towards $f_j^\ast$ in the direction $\vec{e}_j$, $j=0,\hdots, J$, using (\ref{Streaming}).\\

\item[\emph{iii}.] Computation step of the macroscopic variables $\rho$ and $\vec{u}$.\\

\item[\emph{iv}.] Computation step of the microscopic variables $f_j^{eq}$, $j=0,\hdots, J$.\\

\item[\emph{v}.] Collision step, in order to obtain the updated densities $f_j$, $j=0,\hdots, J$, using (\ref{Collision}).\\

\item[\emph{vi}.] Return to steps \emph{ii}. to \emph{v}.

\end{enumerate}

\section{Spurious lattice solitons}
\label{LatticeSol}

\noindent The discrete solution associated with the \emph{LB} numerical scheme will admit spurious
solitary waves, and therefore spurious local energy pile-up and local
sudden growth of the error, if the discrete relation used to implement the scheme is satisfied by a solitary wave.\\

\noindent Following \cite{feng1}, \cite{ClDavid1}, \cite{ClDavid2}, \cite{ClDavid3}, \cite{ClDavid4}, and using \cite{Li}, \cite{whitham}, \cite{ablowitz}, \cite{dodd}, \cite{johnson},
\cite{ince}, \cite{birk}, \cite{Polyanin}, we search solitary waves solution components under the form:

\begin{equation}
\label{Soliton} u_i(\vec{r},t )=\displaystyle \sum_{j=0}^J \left \lbrace U_{ i,j}\,\text{sech}\left [   k_j\,\left (x_i-\left (\vec{e}_j\right)_i\,t\right) \right]
+V_{ i,j}\,\text{tanh}\left [   k_j\,\left (x_i-\left (\vec{e}_j\right)_i\,t\right) \right ] \right \rbrace
\end{equation}

\noindent where

$$\vec{r}= (x_i)_{1\leq i\leq d} \quad , \quad \vec{e}_j= \left ( \left (\vec{e}_j\right)_i \right)_{1\leq i\leq d} \quad ,\quad j=0,\hdots,J$$

\noindent and where $U_{ i,j}$, $V_{ i,j}$, $k_j$, \mbox{$i=1,\hdots,d$}, \mbox{$j=0,\hdots,J$}, are real constants.\\

\noindent The question one may ask is how to switch from the scheme, which is discrete, to this continuous expression ? In so far as if one can find a solution of the above form (\ref{Soliton}) that fits each step of the scheme, is the answer. This is how we will proceed.\\

\noindent Starting from the fact that once the discrete populations $f_j$ are known, the main fluid quantities can be obtained by simple linear summation upon the discrete speeds,
it is legitimate, for those densities, to be of the above form (\ref{Soliton}), for the same reasons as previously.\\

\subsection{A one-dimensional example: the D1Q3 model}

\noindent In the following, we examine a one-dimensional case, given by the flow of an incompressible fluid, in a domain $[0,L]$, $L>0$, obeying the Navier-Stokes equation, in order to test the D1Q3 model, which has  a  zero  velocity  and  two
oppositely directed ones, moving thus the fluid particle to the left and right neighbour lattice sites.\\

\begin{figure}[htbp]
\center{\includegraphics[width=8cm]{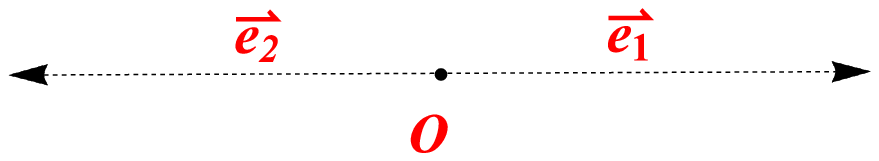}}
\center{\footnotesize{\textbf{\caption{}}}}
\centerline{\footnotesize{\textbf{Discrete velocity vectors of the D1Q3 model}}}
\end{figure}

\noindent The velocities are:

$$\vec{e}_0=0 \quad , \quad \vec{e}_1=1 \quad , \quad \vec{e}_2=-1$$

\noindent while the weights are:

$$  w_{0}=\displaystyle \frac{2}{3}\quad , \quad
w_{ 1}=\displaystyle \frac{1}{6}\quad ,\quad w_{2}=\displaystyle \frac{1}{6} $$

\noindent The conservation of mass quantity writes:
\begin{equation}
\label{D1Q3}
f_1=  \rho\,u +f_2
\end{equation}

\noindent Since the density is a conserved quantity, we can suppose it to be constant, and normalized.  In order to determine wether there exist, or not, solitary waves solutions, one requires to take into account related boundary conditions, in the case of an imposed velocity. Then, using (\ref{Soliton}), we search the densities $f_i$, \mbox{$i=1,2$}, under the form:

$$f_i( y,t)=\displaystyle \sum_{j=0}^2 \left \lbrace \Phi_{i,j}\,\text{sech}\left [  k_j\,\left (x-\left (\vec{e}_j\right)_x\,t\right)  \right  ]
+
\Psi_{i,j}\,\text{tanh}\left [   k_j\,\left (x-\left (\vec{e}_j\right)_x\,t\right) \right ] \right \rbrace$$

 \noindent and the horizontal velocity as:

$$
u ( x,t) = \displaystyle \sum_{j=0}^2 \left \lbrace U_{ j}\,\text{sech}\left [   k_j\,\left (x-\left (\vec{e}_j\right)_x\,t\right) \right]
+V_{ j}\,\text{tanh}\left [   k_j\,\left (x-\left (\vec{e}_j\right)_x\,t\right) \right ] \right \rbrace  $$

\noindent where, for $j=0,\hdots,2$, $k_{j }$, $\Phi_{i,j}$, $\Psi_{i,j}$, $U_{ j}$, $V_j$, are real constants to be determined.    \\

By substituting those latter expressions in (\ref{D1Q3}), one gets an equation of the form:

\footnotesize
$$\displaystyle \sum_{j=0}^2 \left \lbrace {\cal F}  \left(\Phi_{i,j} ,  \Psi_{i,j},U_{x,j} \right)\,\text{sech}\left [  k_j\,\left (y-\left (\vec{e}_j\right)_x\,t\right)  \right]
+{\cal G}_i \left(\Phi_{i,j} ,  \Psi_{i,j},V_{x,j} \right)\,\text{tanh}\left [ k_j\,\left (y-\left (\vec{e}_j\right)_x\,t\right)  \right ] \right \rbrace
=0$$

\normalsize
\noindent where we denote by ${\cal F} $ and ${\cal G} $ functions of $\Phi_{i,j} $,  $\Psi_{i,j}$, $U_{ j}$, $V_j$. By independance of the terms in
sech and tanh, one gets, for $j=0,\hdots, 2$:

$$   {\cal F}  \left(\Phi_{i,j} ,  \Psi_{i,j},U_{ j},V_{ j}\right)\, =0$$

$$  {\cal G}  \left(\Phi_{i,j} ,  \Psi_{i,j},U_{ j},V_{ j}\right)\, =0$$

\noindent i.e. a linear system of two equations, the unknowns of which are the $\Phi_{i,j}$, $\Psi_{i,j}$, $U_{ j}$, $V_{ j}$. It happens that in this D1Q3 model, the system is a very simple one:

$$\Phi_{1,j}=U_j+\Phi_{2,j}\quad , \quad \Psi_{1,j}=V_j+\Psi_{2,j}\quad , \quad j=0,\hdots,2$$

\noindent which admits several sets of solutions. For sake of simplicity,
we have choosen the following one:

$$\Phi_{1,1}=1 \quad , \quad \Phi_{2,1}=2 \quad , \quad U_0=1   \quad , \quad U_1=-1 $$

$$\Phi_{1,2}=2 \quad , \quad \Phi_{2,2}=4     \quad , \quad U_2=-2 $$

$$\Psi_{1,1}=4 \quad , \quad \Phi_{2,1}=5 \quad , \quad V_0=1   \quad , \quad V_1=-1 $$

$$\Psi_{1,2}=6 \quad , \quad \Phi_{2,2}=8     \quad , \quad V_2=-2 $$

\noindent It thus exhibits the existence of lattice solitons, related to the discrete numerical scheme, of the form

\begin{equation}
\label{LatticeSoliton1}
 u_i^n=A_i\,\text{Sech}\left [ k_i\,(i\,h-n\,\left ({\vec{e}_j}\right)_i\,\Delta\,t) \right ]+B_i\,\text{Tanh}\left [ k_i\,(i\,h-n\,\left ({\vec{e}_j}\right)_i\,\Delta\,t) \right ]
\quad , \quad \left (B_i,k_i \right )\,\in \, \R^2
\end{equation}


\vskip 1cm

\noindent Figure 1 displays the related lattice solitary wave, as a function of the mesh points ; $n_x$ denotes the spatial number of mesh points in the $x$-direction, 
$n_t$ the time one.

\vskip 1cm

\begin{figure}[htbp]
\center{\includegraphics[width=8cm]{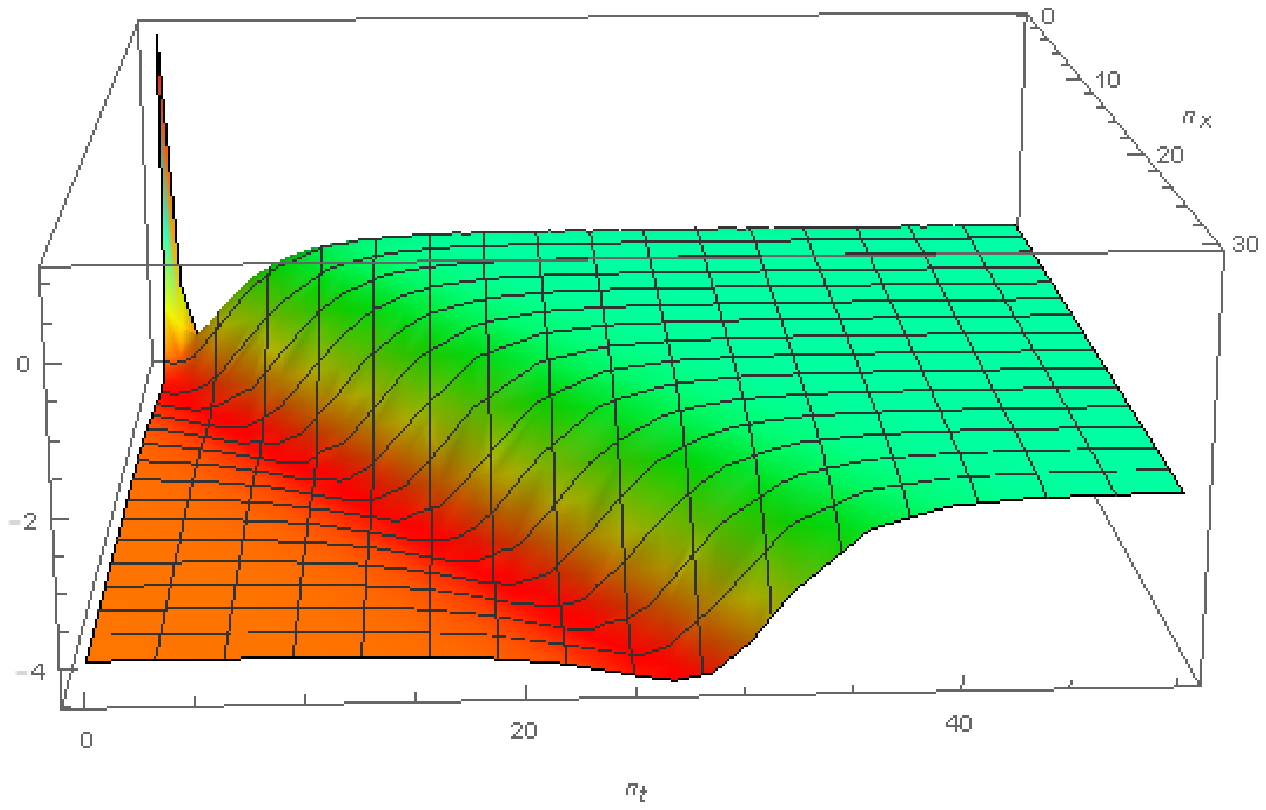}}
\label{LatticeSolitonLB.eps}
\center{\footnotesize{\textbf{\caption{}}}}
\centerline{\footnotesize{\textbf{A "lattice solitary wave",}}}
\centerline{\footnotesize{\textbf{as a function of the mesh points, in the case of the D1Q3 model}}}
\end{figure}

\subsection{A two-dimensional example: The case of an isothermal Poiseuille flow between two parallel plates}

\noindent The case of an isothermal Poiseuille flow, driven by a pressure gradient in the $x$-direction, between two horizontal, parallel, plates is an interesting one: if the flow is a two-dimensional one, yet, since it is a longitudinal one, one only has to deal with the longitudinal velocity component, $u_x$, which happens to depend only on the vertical space variable $y$. In our case, that means that if there exists solitary waves, they also will be horizontal ones, and functions of $y$. Also, since the exact analytical expression of the velocity is known (the flow develops a parabolic velocity profile), results can be tested profitably. In the following, we will denote by $H >0$ the distance between the two plates.\\

\noindent The velocity vectors of the D2Q9 model are:

$$\vec{e}_0=\vec{0}= (0,0)$$

$$\vec{e}_1=(1,0) \quad, \quad \vec{e}_2=(0,1) \quad, \quad \vec{e}_3=(-1,0) \quad, \quad \vec{e}_4=(0,-1) $$

$$\vec{e}_5=(1,1) \quad, \quad \vec{e}_6=(-1,1) \quad, \quad \vec{e}_7=(-1,-1) \quad, \quad \vec{e}_8=(1,-1) $$

\begin{figure}[htbp]
\center{\includegraphics[width=8cm]{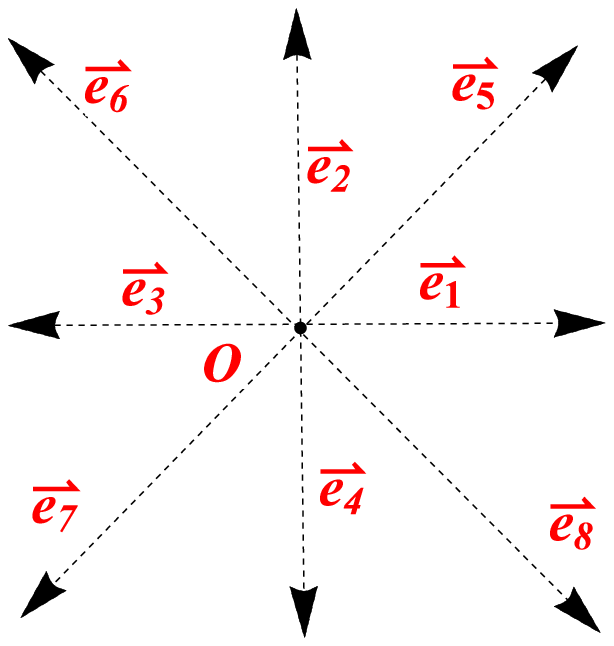}}
\center{\footnotesize{\textbf{\caption{}}}}
\centerline{\footnotesize{\textbf{Discrete velocity vectors of the D2Q9 model}}}
\end{figure}

\noindent In the specific case of the D2Q9 model, one has:

 $$f_j^{eq}=w_j\,\rho \left (1+ \displaystyle \frac{3\,\vec{e}_j\cdot \vec{ u}}{c_0^2}+ \displaystyle \frac{9\,\left(\vec{e}_j\cdot \vec{ u}\right)^2}{2\,c_0^4}-\displaystyle \frac{3\,\vec{u}^2}{2\,c_0^2}\right)$$

 \noindent The weights of the population are given by:

 $$w_0=\displaystyle \frac{1}{9}$$

 $$\forall\,j\,\in\, \left \lbrace 1,2,3,4 \right \rbrace \, :\quad w_j=\displaystyle \frac{1}{9}$$

 $$\forall\,j\,\in\, \left \lbrace 5,6,7,8  \right \rbrace \, :\quad w_j=\displaystyle \frac{1}{36}$$

\noindent As in the above, in order to determine wether there exist, or not, solitary waves solutions, one requires to take into account related boundary conditions, in the case of an imposed velocity. We thus place ourselves in the case of the Zhou-He scheme \cite{ZhouHe}. The boundary between the fluid and the plates overlap the domain nodes. If one chooses to solve the undetermined population of the south boundary, $\vec{e}_2$, $\vec{e}_5$, $\vec{e}_6$, conservation of the mass and quantity of movement yield:

\begin{equation}
\label{Syst1D2Q9}\left \lbrace \begin{array}{ccc}
f_2+f_5+f_6&= &\rho-(f_0+f_1+f_3+f_4+f_7+f_8)\\
f_5-f_6&=&\rho\,u_x-(f_1-f_3-f_7+f_8)\\
f_2+f_5+f_6&=& f_4+f_7+f_8
\end{array}\right.
\end{equation}

\noindent One can then, as in \cite{ZhouHeScheme} use the bounceback of non-equilibrium in the direction normal to the wall:

$$f_2-f_2^{eq}=f_4-f_4^{eq}$$

\noindent which leads to:

\begin{equation}
\label{Syst2D2Q9}\left \lbrace \begin{array}{ccc}
f_2 &=&f_4 \\
f_5 &=&f_7-\displaystyle \frac{f_1-f_3}{2}+ \displaystyle \frac{1}{2}\,\rho\,u_x \\
f_6&=&f_8+\displaystyle \frac{f_1-f_3}{2}- \displaystyle \frac{1}{2}\,\rho\,u_x
\end{array}\right.
\end{equation}

\noindent Let us now study the existence of solitary waves solutions of (\ref{Syst1D2Q9}) and (\ref{Syst2D2Q9}). We assume the density to be constant, and normalized. Then, we search the densities
$f_i$, $i=1,\hdots,8$, and the longitudinal velocity component $u_x$, as functions of the vertical space variable $y$, and of the time one, $t$, under the form:

$$f_i( y,t)=\displaystyle \sum_{j=0}^8 \left \lbrace \Phi_{i,j}\,\text{sech}\left [  k_j\,\left (y-\left (\vec{e}_j\right)_x\,t\right)  \right  ]
+
\Psi_{i,j}\,\text{tanh}\left [   k_j\,\left (y-\left (\vec{e}_j\right)_x\,t\right) \right ] \right \rbrace$$

$$
u_x( y,t) = \displaystyle \sum_{j=0}^8 \left \lbrace U_{x,j}\,\text{sech}\left [   k_j\,\left (y-\left (\vec{e}_j\right)_x\,t\right)\right ]
+V_{x,j}\,\text{tanh}\left [   k_j\,\left (y-\left (\vec{e}_j\right)_x\,t\right) \right ] \right \rbrace  $$

\noindent where, for $j=0,\hdots,8$, $k_{j }$, $\Phi_{i,j}$, $\Psi_{i,j}$, $U_{x,j}$, $V_{x,j}$, are real constants to be determined.    \\

\noindent The boundary conditions on the plates require:

\begin{equation}
\label{CLD2Q9}\left \lbrace \begin{array}{ccccc}
u_x( 0,t)&=&\displaystyle \sum_{j=0}^8 \left \lbrace U_{x,j}\,\text{sech}\left [   k_j\,\left ( -\left (\vec{e}_j\right)_x\,t\right)\right ]
+V_{x,j}\,\text{tanh}\left [   k_j\,\left ( -\left (\vec{e}_j\right)_x\,t\right) \right ] \right \rbrace&=&0 \\
u_x( H,t)&=&\displaystyle \sum_{j=0}^8 \left \lbrace U_{x,j}\,\text{sech}\left [   k_j\,\left (H-\left (\vec{e}_j\right)_x\,t\right)\right ]
+V_{x,j}\,\text{tanh}\left [   k_j\,\left (H-\left (\vec{e}_j\right)_x\,t\right) \right ] \right \rbrace&=&0 \\
\end{array}
\right.
\end{equation}

\noindent By substituting those latter expressions in (\ref{Syst1D2Q9}) and (\ref{Syst2D2Q9}), one gets a system of six equations of the form:

\footnotesize
$$\displaystyle \sum_{j=0}^8 \left \lbrace {\cal F}_i \left(\Phi_{i,j} ,  \Psi_{i,j},U_{x,j} \right)\,\text{sech}\left [  k_j\,\left (y-\left (\vec{e}_j\right)_x\,t\right)  \right]
+{\cal G}_i \left(\Phi_{i,j} ,  \Psi_{i,j},V_{x,j} \right)\,\text{tanh}\left [ k_j\,\left (y-\left (\vec{e}_j\right)_x\,t\right)  \right ] \right \rbrace
=0$$

\normalsize
\noindent where, for $i=1,\hdots, 6$, we denote by ${\cal F}_i$ and ${\cal G}_i$ functions of $\Phi_{i,j} $,  $\Psi_{i,j}$, $U_{x,j}$, $V_{x,j}$. By independance of the terms in
sech and tanh, one gets, for $i=1,\hdots, 6$:

$$   {\cal F}_i \left(\Phi_{i,j} ,  \Psi_{i,j},U_{x,j},U_{y,j}\right)\, =0$$

$$  {\cal G}_i \left(\Phi_{i,j} ,  \Psi_{i,j},U_{x,j},U_{y,j}\right)\, =0$$

\noindent i.e. a linear system in 12 equations, the unknowns of which are the $\Phi_{i,j}$, $\Psi_{i,j}$, $U_{x,j}$, $V_{x,j}$. \\

\noindent Taking into account the boundary conditions (\ref{CLD2Q9}), resolution with a symbolic calculus tool (Mathematica for instance) shows that this system admits several sets of solutions. For sake of simplicity,
we have choosen the following one:

$$U_{x,0}=-U_{x,1}-U_{x,2}-U_{x,3}-U_{x,4}-U_{x,5}-U_{x,6}-U_{x,7}-U_{x,8}$$

$$
U_{x,1}= 1 \quad, \quad U_{x,2}=-1 \quad, \quad U_{x,3} = 1\quad, \quad U_{x,4} = 1$$

$$ U_{x,5} = 1\quad, \quad U_{x,6} = 6\quad, \quad U_{x,7}= 1 \quad, \quad U_{x,8} = 5$$

$$V_{x,1} = 1\quad,V_{x,2} = 1\quad, \quad V_{x,3}  = 3\quad, \quad V_{x,4}  = 4$$

$$ V_{x,5}  = 5\quad, \quad V_{x,6}  = 6\quad, \quad V_{x,7}  = 7\quad, \quad V_{x,8}  = 8$$

$$V_{x0}=-\left \lbrace  U_{x,j} \,\text{sech}(k_j\,H)+ \displaystyle \sum_{j=1}^8 \left \lbrace   U_{x,j} \,\text{sech}(k_j\,H)+V_{x,j} \,\text{tanh}\, (k_j\,H )
 \right \rbrace \right \rbrace \,\text{cotanh}\, (k_0\,H)$$


$$k_0 = 1 \quad, \quad  k_1 = 0.1 \quad, \quad  k_2 = 10 \quad, \quad  k_3 = 4$$

$$ k_4 = 5 \quad, \quad  k_5 = 6 \quad, \quad  k_6 = 7 \quad, \quad  k_7 = 8 \quad, \quad  k_8 = 9$$

\noindent Figure 2 displays the related lattice solitary wave, as a function of the normalized vertical coordinate, which can be compared to the exact analytical solution.

\vskip 1cm

\begin{figure}[htbp]
\center{\includegraphics[width=8cm]{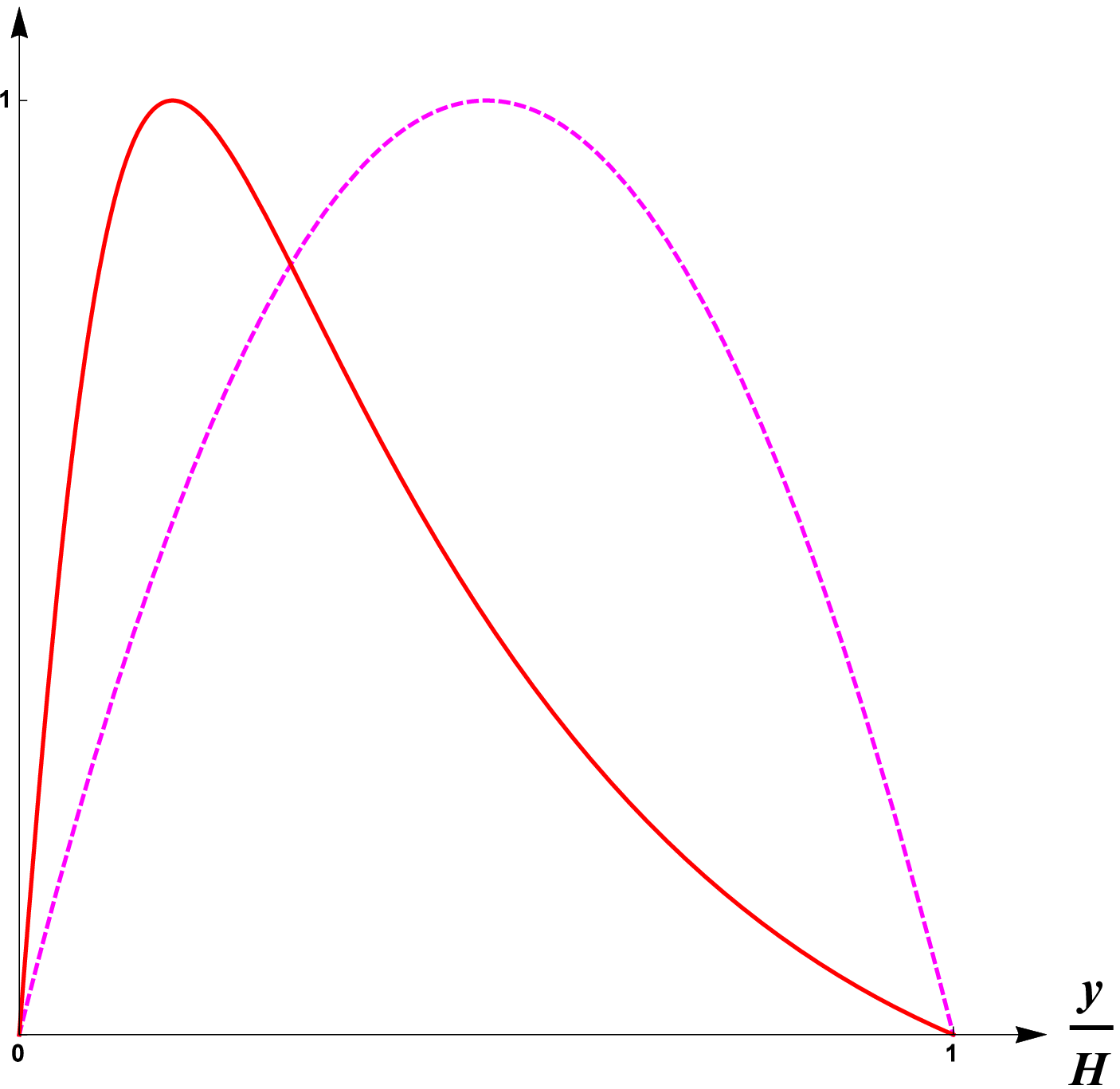}}
\label{LatticeSolitonLB.eps}
\center{\footnotesize{\textbf{\caption{}}}}
\centerline{\footnotesize{\textbf{A "lattice solitary wave",}}}
\centerline{\footnotesize{\textbf{in the case of a Poiseuille flow between two parallel plates (in red),}}}
\centerline{\footnotesize{\textbf{compared with the analytical exact solution (dashed plot).}}}
\end{figure}

\section{Concluding remarks}

The existence of spurious numerical lattice solitary waves for Lattice Boltzmann schemes has been proved. Such lattice solitary waves, which are not solutions of the exact continuous original equation, nevertheless satisfy the numerical scheme, appearing as parasitic solutions of the correct one.
Such schemes will be referred to as structurally instable ones. A solution to avoid such spurious solitary waves, would be to "lock" the scheme by adding a test in the main loop.\\
Such spurious solitary waves have constant energy, and therefore the numerical error norm does not vanish at arbitrary long integration times on unbounded numerical domains.\\

One may ask why do such solutions occur ? A partial answer can be found in the equivalent continuous equation, the principle of which is the following: for a small time step $\Delta t$, and the associated space scale $h$, a Taylor expansion in $\Delta t$ leads to establishing equivalent continuous equations as a formal limit. This matter has been explored by F. Dubois in \cite{Dubois}, where it is proved that, at the first order in $\Delta t$, the equivalent equation is the Euler equations, which are known to admit solitary waves solutions (see, for instance, \cite{}).


\begin{thebibliography}{99}




\bibitem {ClDavid1} Cl. David, P. Sagaut, {Structural stability of discontinuous Galerkin schemes}, Acta Applicandae, under press.



\bibitem {ClDavid2} Cl. David, P. Sagaut, {Structural stability of finite dispersion-relation preserving
schemes}, Chaos, Solitons and Fractals, \textbf{41} (4), 2009, 2193-2199.


\bibitem {ClDavid3} Cl. David, P. Sagaut, {Spurious solitons and structural stability of finite
difference schemes for nonlinear wave equations}, Chaos, Solitons and Fractals, \textbf{41} (2), 2009, 655-660.


\bibitem {ClDavid4} Cl. David, R. Fernando, Z. Feng, {A note on "general solitary wave solutions
of the Compound Burgers-Korteweg-de Vries Equation"}, Physica A:
Statistical and Theoretical Physics, \textbf{375} (1), 2007, 44-50.

\bibitem {Hardy} J. Hardy, Y. Pomeau and O. de Pazzis, Time Evolution of a Two-Dimensional
Classical Lattice System, Physical Review Letters, \textbf{31}, 1973, 276-279.

\bibitem {Humieres1} D. d'Humi\`eres, P. Lallemand and U. Frisch, Lattice gas models for 3D-hydrodynamics,
Europhysics Letters, \textbf{2}(4), 1986, 291-297.

\bibitem {Frisch} U. Frisch, B. Hasslacher and Y. Pomeau, Lattice gas automata for the Navier–Stokes
equation", Physical Review Letters, \textbf{56}(14), 1986, 1505-1508, 1986.


\bibitem {Higuera1} F. Higuera, J. Jimenez, and S. Succi. Boltzmann approach to lattice gas
simulations, Europhys. Lett, \textbf{9}, 1989, 663.


\bibitem {Higuera2}  F. Higuera, J. Jimenez, and S. Succi. Lattice Gas dynamics with enhanced
collision. Europhys. Lett, 9,1989, 345.

\bibitem {Bhatnagar} P. Bhatnagar, E. Gross and M. Krook. "A Model for Collision Processes in Gases.
I. Small Amplitude Processes in Charged and Neutral One-Component Systems",
Physical Review, \textbf{94}, 1954, 511-525.

\bibitem {Qian} Y. Qian, D. d'Humi\`eres, and P. Lallemand, Lattice BGK models for Navier-
Stokes equation, Europhys. Lett,\textbf{ 17}, 1992, 470–84.

\bibitem {Humieres} D. d'Humi\`eres, Generalized Lattice-Boltzmann Equations", in Rarefied Gas Dynamics:
Theory and Simulations, AIAA Progress in Astronautics and
Astronautics, \textbf{1959}, 1992, 450-458.


\bibitem {Chopard1} B. Chopard, A. Dupuis, A. Masselot, and P. Luthi. Cellular Automata and
Lattice Boltzmann techniques: An approach to model and simulate complex
systems, Advances in Complex Systems, \textbf{5}, 2002, 103–246.


\bibitem {Chopard2} B. Chopard and M. Droz, Cellular Automata Modeling of Physical
Systems, Cambridge University Press, 1998.

\bibitem {Reider} M. B. Reider and J. D. Sterling, Accuracy of Discrete-Velocity BGK modor the simulation of the incompressible Navier-Stokes equations. Comput.
fluids, \textbf{24}(4), 1995, 459–467.

\bibitem {Lallemand} P. Lallemand, L.-S. Luo, Theory of the lattice Boltzmann method: Dispersion,
dissipation, isotropy, Galilean invariance, and stability, Physical Review E, \textbf{61},
2000, 6546-6562.

\bibitem {shan} Shan, Xiaowen and Yuan, Xue-Feng and Chen, Hudong,
    {Kinetic theory representation of hydrodynamics: a way beyond the Navier–Stokes equation},
      {J. Fluid Mech.},
      {550},  {2006},
    413-441.





\bibitem{feng1} Z. Feng, G. Chen, Solitary Wave Solutions of the Compound
Burgers-Korteweg-de Vries Equation, {Physica A}, \textbf{352}, 2005,
419-435.




\bibitem{Li} B. Li, Y. Chen Y. and H.Q. Zhang, Explicit exact solutions for new general
two-dimensional KdV-type and two-dimensional KdV–Burgers-type
equations with nonlinear terms of any order, {J. Phys. A
(Math. Gen.)}, \textbf{35}, 2002, 8253–8265.


\bibitem{whitham} G. B. Whitham, Linear and Nonlinear Waves,
Wiley-Interscience, New York, 1974.

\bibitem{ablowitz} M. J. Ablowitz, H. Segur, Solitons and the Inverse Scattering
Transform, SIAM, Philadelphia, 1981.


\bibitem{dodd} R. K. Dodd, J.C. Eilbeck, J. D. Gibbon, H.C. Morris, Solitons and Nonlinear Wave
Equations, London Academic Press, London, 1983.

\bibitem{johnson} R.S. Johnson, A Modern Introduction to the Mathematical Theory of
Water Waves, Cambridge University Press, Cambridge, 1997.

\bibitem{ince} E.L. Ince, Ordinary Differential Equations, Dover
Publications, New York, 1956.




\bibitem{birk} G. Birkhoff, G.C. Rota, Ordinary Differential
Equations, Wiley, New York, 1989.


\bibitem{Polyanin} A.D. Polyanin, V. F. Zaitsev, Handbook of Nonlinear Partial Differential
Equations, Chapman and Hall/CRC, 2004.


\bibitem{ZhouHe} Q. Zou, and X. He, Pressure and velocity boundary conditions for the lattice Boltzmann, J. Phys.
Fluids, \textbf{9}, 1997, 1591-1598.



\bibitem{Dubois} F. Dubois, Equivalent partial differential equations of a lattice Boltzmann scheme, Computers and Mathematics with Applications, \textbf{55}, 2008, 1441–1449.




\end{thebibliography}
\end{document}